\documentclass[useAMS]{mn2e}

\usepackage{graphicx}

\title[]{}

\author[]{}

\title[Probabilistic $M/L$ values for ultra-faint dSphs] {Probabilistic
distributions of $M/L$ values for ultra-faint dwarf spheroidal galaxies:
stochastic samplings of the IMF} 
\author[X. Hernandez] {X. Hernandez  \\ Instituto de
  Astronom\'{\i}a, Universidad Nacional Aut\'{o}noma de M\'{e}xico,
  Apartado Postal 70--264 C.P. 04510 M\'exico D.F. M\'exico. \\}

\date{Released 2010 Xxxxx XX}

\pagerange{\pageref{firstpage}--\pageref{lastpage}} \pubyear{2009}

\def\LaTeX{L\kern-.36em\raise.3ex\hbox{a}\kern-.15em

    T\kern-.1667em\lower.7ex\hbox{E}\kern-.125emX}

\begin{document}

\label{firstpage}

\maketitle

\begin{abstract}
We explore the ranges and distributions which will result for the intrinsic stellar Mass to Light ratios
($M/L$) of single stellar populations, at fixed IMF, age and metallicity, from the discrete stochastic
sampling of a probabilistic IMF. As the total mass of a certain stellar population tends to infinity, 
the corresponding $M/L$ values quickly converge to fixed numbers associated to the particulars
of the IMF, age, metallicity and star formation histories in question. When going to small
stellar populations however, a natural inherent spread will appear for the $M/L$ values, which will
become probabilistic quantities. For the recently discovered ultra-faint local dwarf spheroidal
galaxies, with total luminosities dropping below $10^{3} L_{V}/L_{\odot}$, it is important to asses
the amplitude of the probabilistic spread in inherent $M/L$ values mentioned above. The total baryonic 
masses of these systems are usually estimated from their observed luminosities, and the assumption of a
fixed, deterministic $M/L$ value, suitable for the infinite population limit of the
assumed ages and metallicities of the stellar populations
in question. This total baryonic masses are crucial for the testing and calibrating of structure formation
scenarios, as the local ultra-faint dwarf spheroidals represent the most extreme galactic scales
known. Also subject to reliable $M/L$ values, is the use of these systems as possible discriminants 
between dark matter and modified gravity theories. By simulating large collections of stellar populations,
each consisting of a particular collection of individual stars, we compute statistical distributions for the
resulting $M/L$ values. We find that for total numbers of stars in the range of what is observed for the local
ultra-faint dwarf spheroidals, the inherent $M/L$ values of stellar populations can be expected to vary
by factors of upwards of 3, interestingly, systematically skewed towards higher values than what is obtained
for the corresponding infinite stellar population limit $M/L$. This can serve to explain part of the spread
in reported baryonic masses for these systems, which also appear as shifted systematically towards high
dark to baryonic mass ratios at fixed stellar velocity dispersions, when going to the ultra-faint limit.

\end{abstract}


\section{Introduction}

In determining the inherent mass to light, $M/L$, ratios of stellar populations, it is 
customary to treat the distribution of stellar masses through a probability function, the IMF.
When dealing with a large galaxy, or a large stellar population in general, it is 
justified to think of the IMF as a densely sampled probability function. In practice, this
translates into models where the IMF appears as a series of weighting factors
to be applied to the heavily mass dependent stellar luminosities, when calculating the overall
$M/L$ ratios for stellar populations. 

In going to small star formation events, individual star forming regions, fractions of
small galaxies or the stellar populations of the local ultra-faint dwarf spheroidals recently available to detailed
observation however, the standard assumption of stellar populations, that the IMF 
is a densely sampled probabilistic distribution function, breaks down. If we are dealing 
with a small star formation event, calculating the resulting stellar $M/L$ will require
the explicit inclusion of the probabilistic nature of the IMF, in a regime where this function
is being only poorly and discretely sampled. 

The consequences of this change in regime are amplified by the strong power law
character of the IMF, further compounded by the heavily mass weighted dependence of  
stellar luminosities. In a sample of large stellar populations of fixed total mass, the actual number of 
giant stars for any one will vary by only a fraction of a percent, due to the 
intrinsic variance associated with the probabilistic sampling of the IMF. The resulting 
variance in the intrinsic $M/L$ values will be correspondingly small, and is hence customarily
ignored. If the total mass of stars is of up to a few thousand $M_{\odot}$ however,
the intrinsic probabilistic variance of a standard IMF will lead one to expect only  
a few giant stars. It is clear that the resulting intrinsic variance to be expected
in the final $M/L$ values resulting from such a population will be of a large factor, 
with distributions heavily dependent on both the total mass and age e.g. Cervi\~no \& Luridiana (2006),
Carigi \& Hernandez (2008). Both the stochastic effects of IMF sampling for very low total mass
populations, and correlations between intrinsic IMF and total star formation masses have been
shown to be important by Weidner \& Kroupa (2004), Weidner \& Kroupa (2006) and Koeppen et al. (2007),
for star formation episodes resulting in a few thousand stars, with of order of $10^{5} M_{\odot}$ in 
total gas mass involved, considering typical efficiency factors of a few percent.

The statistical variance in the light output of a stellar population has been
studied previously e.g. Tonry \& Schneider (1988), Cervi\~{n}o \& Valls-Gabaud (2003). The most important practical application
of which has been the development of the surface density fluctuation
method of distance determination in galaxies, Tonry \& Schneider (1988). Similarly, such studies
lead to the realization that the intrinsic variances in populations
of even thousands of stars, in certain observed bands or emission lines, can lead to relevant effects.
The chemical consequences of an stochastic star formation event have recently been studied by 
Koeppen et al. (2007). Also, Carigi \& Hernandez (2008) showed that the intrinsic variance present in 
populations of even more than a few thousand stars, 
can be significant towards explaining part of the scatter in abundance ratios observed
within the classical local dwarf spheroidals. Similarly, Cescutti (2008) applied precisely such ideas 
to explain the spread in neutron capture elements for low metalicity stars in the Solar Neighbourhood.
Using results of population synthesis codes which assume the IMF has been densely
sampled, and which yield a unique answer for certain observed properties of a stellar population
of fixed mass and metallicity, can lead to the over-interpretation of observations if one is 
not careful e.g., Cervi\~{n}o \& Valls-Gabaud (2003).

The recently discovered and studied ultra-faint local dwarf spheroidal galaxies (e.g. Belokurov et al. 2006,
Simon \& Geha 2007, Geha et al. 2009, Martin et al. 2009), with total luminosities of the order of $10^{3}$ 
in solar units and below, and total dynamical $M/L$ values $>10^{3}$, represent a population of objects 
where the intrinsic variance in the stellar $M/L$ ratios, at fixed age and metallicity, might become important. 
To date, the inferences of total baryonic masses for these objects have been performed in the standard way, 
through the use of $M/L$ ratios thought of as deterministic parameters, as appropriate for infinite stellar 
populations, e.g. Angus (2008) or McGaugh \& Wolf (2010). These total baryonic masses become important in 
determining the baryonic to dark matter ratios of the systems in question (e.g. Sanchez-Salcedo  et al. 2006 
or Walker et al. 2009), the calibrating and testing of structure formation models at the smallest galactic 
scales (e.g. Strigari et al. 2010), or the testing of modified gravity theories, where gravity is assumed to 
couple exclusively to the observable baryonic mass e.g. Hernandez et al. (2010), Iorio (2010). Given however 
the negligible dynamical relevance of the stellar components (under standard gravity), the dark matter to 
total mass ratios would remain unchanged at values very close to 1.

In this paper we calculate the range in stellar $M/L$ values for small stellar populations of fixed input parameters, 
which result from the intrinsic statistical variations of a discretely and poorly sampled standard IMF. 
Throughout this paper we shall use $M/L$ to denote total intrinsic stellar mass to light ratios, and not
in connection with dynamical mass to light ratios, which of course are completely robust to the
considerations explored here. As we shall see, the low mass weighted character of the IMF will result in a 
slow drift in the mean $M/L$ values of stellar populations, as the total stellar mass goes down. 
More important will be the appearance of a large spread in the resulting $M/L$ values, which will shift 
qualitatively from being the deterministic proportionality factors which they are for large stellar 
populations, to becoming probabilistic entities with broad distributions skewed towards high values. 
As the total mass diminishes the stochastic effects on the $M/L$ ratios appropriate for star formation 
events will increase. Therefore, it is the regime of the ultra-faint dwarf spheroidals, very small 
systems with old ages, where the effects being explored will be largest.

We simulate stellar populations having various total masses by sampling directly an assumed IMF, 
this produces a discrete collection of stars, which is then used together with a detailed isochrone library
to produce distributions of $M/L$ values for collections of stellar populations having various fixed total
masses. We explore the resulting distributions for $M/L$ ratios as functions of the 
metallicity, stellar age and the total mass, by directly keeping track 
of each individual star formed. Repeating the process a large number of times 
yields a distribution of $M/L$ values resulting from the same input parameters.

The paper is organized as follows: The construction of discrete IMF realizations and their use 
in constructing stochastic distributions of $M/L$ values for simple stellar populations of fixed input parameters 
is described in section (2), section (3) gives our results for total stellar masses in the ranges of the observed
ultra-faint dwarf spheroidals, and section (4) presents our conclusions.

\section{Constructing Statistical $M/L$ values}

In order to explore the range of intrinsic M/L values which small stellar
populations will present, we begin by setting up a discrete IMF. We assume
a fixed underlying probabilistic IMF, from which stochastic samplings will be constructed, 
discrete collections of individual stars. We take the IMF of Larson (1998):

\begin{equation}
dN/dlog \hspace{2pt} m \propto (1+m/m_{s})^{-1.35},
\end{equation}

\noindent where a choice of $m_{s}=0.4 M_{\odot}$ adequately serves to reproduce a present day
solar neighbourhood IMF, with a mean mass close to 1 $M_{\odot}$, e.g. Hernandez \& Ferrara (2001). 
We have taken lower and upper mass bounds of 0.09 $M_{\odot}$ and 20 $M_{\odot}$ for the above IMF, fixed
throughout this study.
Although the details of our study will be slightly sensitive to the choice of this function, the 
trends we describe and our conclusions are generic to any IMF found in the literature, where the probability
of picking a certain mass strongly decreases with stellar mass. These discrete IMFs will only tend to
the infinite mass limit for very large total stellar masses. Even for total stellar masses of a few
thousand $M_{\odot}$, which one could naively assume to constitute 'statistical samples', the distribution
of stars above $1 M_{\odot}$ will be dominated by shot noise effects, e.g. see Carigi \& Hernandez (2008).
This leads to an effective upper stellar mass which decreases as the total mass of a stellar population
goes down e.g. Massey (1998), Weidner \& Kroupa (2004).
This in turn leads to substantial scope for variations in the resulting intrinsic $M/L$ values, as the light
output of a stellar population is heavily dominated by the giants, while the total mass is a much more
robust quantity, anchored on the integral of the main sequence. We start by picking a value
for the total stellar mass of a single stellar population, and then proceed to randomly pick stars
out of the fixed underlying probabilistic IMF, until the chosen total stellar mass has been reached.
This produces a collection of individual stars, a particular discrete IMF.

Next, we use an extensive isochrone library which carefully interpolates directly on the output of stellar
evolutionary codes at fixed stellar phase, having close to 250 masses between a lower mass bound of $0.15 M_{\odot}$
and the tip of the RGB. This was prepared using the stellar evolutionary codes of the Padova group (Girardi et al. 2002),
for use in the probabilistic parameter inference study for globular clusters of 
Hernandez \& Valls-Gabaud (2008). With this at hand, we then assign to each of the individual stars
selected its corresponding $M_{V}$ value. By then adding the corresponding luminosities, we calculate the total
$V$ band luminosity of a particular realisation of the fixed underlying IMF, at a given age and metallicity. 

We have set to zero the luminosity of all stars outside of the mass range of the isochrones. This introduces a slight
error, but one which will not affect our results significantly, as the integrated luminosity of stars between our
lower IMF limit of $0.09 M_{\odot}$ and our lower isochrone mass limit of $0.15 M_{\odot}$ is only a very minor
contribution to the total light output of a stellar population. Beyond our upper isochrone limit at the tip of
the RGB, the number of bright sources is small, and its exclusion does not significantly alter the total light
budget. A further small error is introduced by having also ignored stellar mass loss throughout, i.e.,
the initial mass of the total stellar population is divided by the total V band luminosity within the isochrone
range, to obtain the final $M/L$ value of a particular IMF realization, at a given total mass, age and metallicity.

Depending on the chosen value for the total stellar mass, and on the assigned age of the stellar population, the 
number of stars selected varies from a few hundred to upwards of 70,000, for the range of models
presented here, for each individual discrete IMF realization. Each of these stars is then assigned a $V$ band luminosity,
as described above. The whole process is then repeated 2000 times, changing the random seed of the simulation, 
to construct a distribution of 2000 $M/L$ values, for each fixed age, metallicity and total stellar mass value
presented here.

\section{Resulting Distributions of $M/L$ Values}

\begin{figure}
\includegraphics[angle=0,scale=0.4]{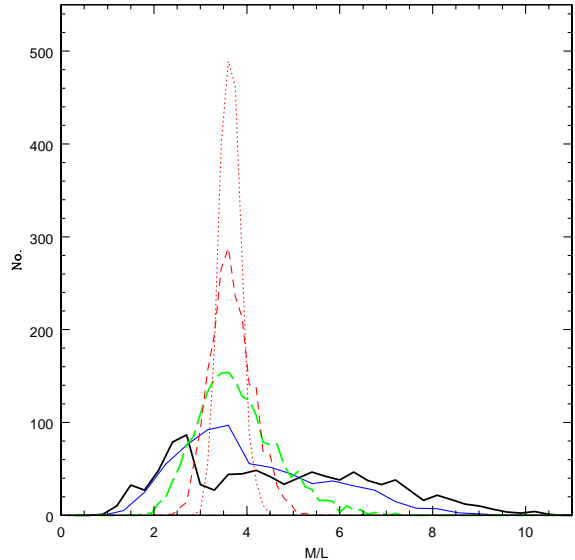}
\caption{Distributions of $M/L$ values for 2000 statistical realizations of single stellar populations,
all at a fixed age of 10.5 Gyr, all at a fixed metallicity of [Z]=-3.0, for different total stellar
masses of $4\times 10^{4}$, $1\times 10^{4}$,  $3\times 10^{3}$, $1\times 10^{3}$, and $5\times 10^{2}$
solar masses, from the most to the lease peaked respectively.}
\label{fig:picture}
\end{figure}

Here we present results for the experiment described in the previous section. We begin by taking
a fixed age of 10.5 Gyr and a fixed metallicity of [Z]=-3 in solar units, parameters as appropriate for 
the old local dwarf spheroidals e.g. Hernandez et al. (2000), Kirby et al. (2008), Sand et al. (2009), Geha et al. (2009), Sand et al.(2010). 
We then simulate 2000 single stellar populations having always a fixed total mass of $5 \times 10^{2} M_{\odot}$. Each results
in a discrete collection of between 650 and 850 stars, all of which are individually assigned a luminosity
from the appropriate isochrone.

A binned distribution for the resulting $M/L$ values for this case is given by the thick solid curve in figure (1). 
A very broad distribution presenting evident fluctuations is apparent, extending to values
of $M/L =11$. This is natural if one considers that the small samples of only several hundred stars from which 
these $M/L$ values have been calculated, reproduce the underlying probabilistic IMF only for low mass stars,
shot noise dominates the distribution already for masses above $1 M_{\odot}$. These low mass stars in turn, typically
have very low luminosities, hence resulting preferentially in $M/L$ values higher than the infinite population
limit for the parameters used, of slightly below 3.5. Occasionally, an IMF realization with over average 
numbers of comparatively massive stars appears, resulting in the extension seen towards lower than average $M/L$ values. 

As we increase the total mass of 
this stellar population to $1 \times 10^{3} M_{\odot}$, we obtain the thin solid curve for the corresponding
distribution of intrinsic $M/L$ values, still at the same fixed age, metallicity and IMF. In this case, we see that
the fluctuations start to disappear, as a smoother distribution results. Still, the thin solid curve of this
case shows a wide range of intrinsic $M/L$ values for the stellar populations, at the fixed parameters being modelled.
The distribution remains skewed towards higher than average values, for the same reasons as mentioned above. We can see
that at these numbers of stars, we can still expect to find $M/L$ values ranging from 2 to 7, even though $1 \times 10^{3}$
might ordinarily be considered an 'statistical number' of stars. As it also happens when considering the intrinsic spread
in the chemical enrichment properties of a stellar population (Carigi \& Hernandez 2008), the heavily low mass
weighted nature of the IMF, plus the strongly top heavy nature of the light output or enrichment properties of stars, 
allows for wide distributions in the intrinsic properties of stellar populations, at fixed input parameters.
 
In going to total stellar masses of  $3 \times 10^{3} M_{\odot}$,  $1 \times 10^{4} M_{\odot}$, and  $4 \times 10^{4} M_{\odot}$,
we obtain the corresponding distributions of intrinsic $M/L$ values, given by the long dashed, short dashed and dotted
curves in figure (1). We see that these distributions tend towards the infinite population limit, $(M/L)_{\infty}$ with mean values which
clearly converge quite rapidly. The inherent spread of the distributions however, takes longer to tend to zero, and even for
quite large numbers of stars of upwards of 60,000 contained in each of the 2000 simulations with results given by the
dotted curve, a noticeable width to the distribution is still evident.

\begin{figure}
\includegraphics[angle=0,scale=0.4]{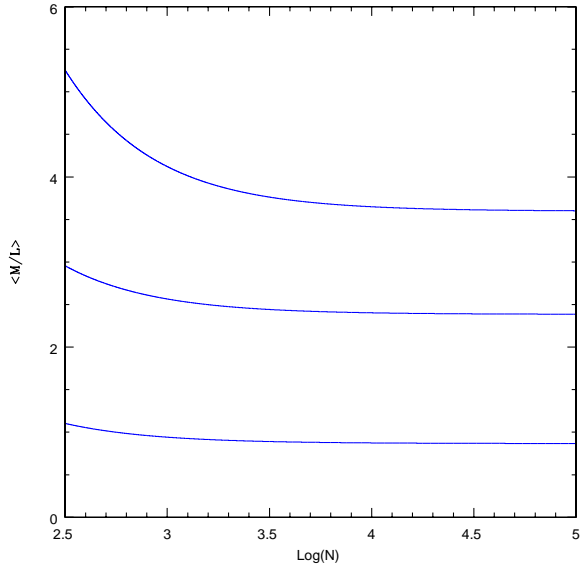}
\caption{Average values of the M/L ratios for statistical realizations of single stellar populations, 
as a function of total stellar mass in solar units, all at a fixed metallicity of [Z]=-3.0, for three
different stellar ages of 10.5 Gyr, 5 Gyr and 2 Gyr, from top to bottom, respectively.}
\label{fig:picture}
\end{figure}

\begin{figure}
\includegraphics[angle=0,scale=0.4]{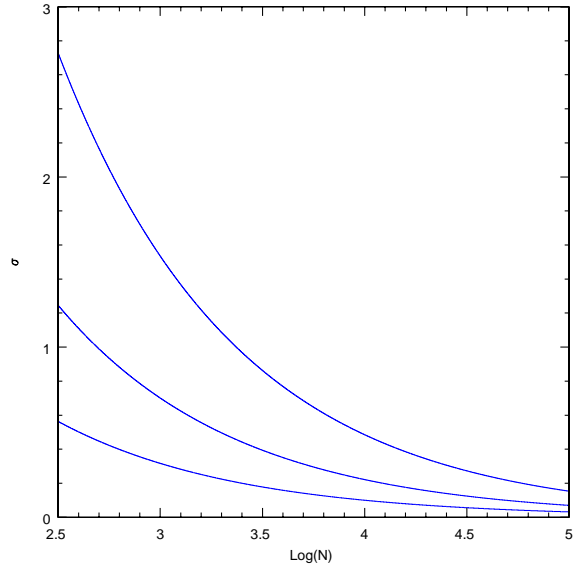}
\caption{Standard deviation for the distributions of M/L ratios for statistical realizations of single stellar 
populations, as a function of total stellar mass in solar units, all at a fixed metallicity of [Z]=-3.0, for three
different stellar ages of 10.5 Gyr, 5 Gyr and 2 Gyr, from top to bottom, respectively.}
\label{fig:picture}
\end{figure}

To better appreciate the manner in which the resulting distributions tend to the infinite population limit, we present figure (2), 
where the average values of the $M/L$ ratios, $<M/L>$, for stochastic realizations of the IMF are given, as a function of the total
stellar mass, for three different stellar ages. The curves correspond to stellar ages of 10.5 Gyr, 5 Gyr and 2 Gyr, from top to
bottom, respectively. As seen from figure (1), the mean values for the distributions of $M/L$ ratios quickly tend to the infinite
population limit, $(M/L)_{\infty}$, also as expected, this occurs at smaller total stellar masses for the younger populations. The drift
towards larger $M/L$ values as the total stellar mass goes down, for the mean of the distributions is also evident. This
effect is fairly important in the case of the oldest populations, and probably accounts for at least part
of the trends seen for the dark to baryonic matter ratios to increase in the ultra-faint dwarf spheroidals. Corresponding plots
at different metallicities are practically indistinguishable, until one reaches metallicities of upwards of $[Z]=-1$, values
no longer relevant for the local dwarf spheroidal galaxies. 

The corresponding plot for the standard deviations, $\sigma$, of the distributions of $M/L$ ratios is given in figure (3).
The three curves correspond to the same three cases of figure (2), again, ages of 10.5 Gyr, 5 Gyr and 2 Gyr, from top to
bottom, respectively. This time we see that although the dispersion in the distribution of $M/L$ values goes down 
as the total mass of the stellar populations increases, this happens at a much slower rate than what characterises the
trend for $<M/L>$ with total stellar masses. We see that for the oldest age, even for stellar populations well into the 
thousands of $M_{\odot}$, dispersions of more than 1 are to be expected. Since the corresponding $<M/L>$ values are of 
around 4, variations in $M/L$ by factors of 2 and above will be frequent. Also, notice that since the distributions
are heavily skewed, the mean and the dispersion offer only crude approximate descriptions, with higher than average values
being the norm.

The resulting  $<M/L> - (M/L)_{\infty}$ and $\sigma$ values can be very accurately described by the following fitting functions:

\begin{equation}
<M/L> -(M/L)_{\infty} = A M_{tot}^{-1}, \;\; \;\;\;\;\;\;\; \sigma = B M_{tot}^{-1/2}, 
\end{equation}

\noindent with the constants $(A, B)$ in eq. (2) having values $(524.72, 48.53)$, $(180.56, 22.16)$, and $(74.84, 10.02)$ for stellar
ages of 10.5 Gyr, 5 Gyr and 2 Gyr, respectively, for metallicities below $[Z]=-1$ in solar units. In the above equations
$M_{tot}$ is the fixed total stellar mass of a set of IMF realizations.

Finally, we present in figure (4) an estimate of the factor by which one can expect the intrinsic $M/L$ values of 
small stellar populations to vary, for metallicities and ages suitable for the local ultra-faint dwarf spheroidals.
The factor $F$ is defined as $(<M/L>+1.5\sigma)/(<M/L>-1.5\sigma)$, to better account for the strongly skewed
distributions present. We see again this factors tending rapidly to $F=1$ as one exceeds total stellar masses
of $3 \times 10^{4} M_{\odot}$, but reaching quite high values of more than 6 for total stellar masses below
$3 \times 10^{3} M_{\odot}$, in the range of parameters inferred for the systems in question. For comparison,
the following of the local ultra-faint dSphs discovered to date have inferred total stellar masses of order 
$10^{3} M_{\odot}$ and below: Bo\"{o}tes II, Ursa Major II, Willman I, and Coma Berenices, with Segue I coming in 
at a mere 600 $M_{\odot}$ in stars, as reported in the compilation of Misgeld \& Hilker (2011).

As can be seen from figure (4),
the differences between the results for ages of 10.5 Gyr and 5 Gyr are much smaller than between results for 5 Gyr and
2 Gyr. In fact, the inherent convergence of the isochrones at large ages implies that results for any stellar ages beyond 10.5 Gyr,
will be scarcely distinguishable from the curves shown for 10.5 Gyr. Thus, the results given for 10.5 Gyr ages are suitable for
the directly inferred ages of some of these systems, e.g. the values of between 12 and 13 Gyr obtained by Sand et al. (2009) 
for the Hercules system, or by Sand et al. (2010) for Leo IV. Also, although the trends presented
will be qualitatively the same for $M/L$ ratios in other bands, the amplitude of the effect presented will grow
towards bluer bands, and decrease towards redder ones, as the relative contribution of the different evolutionary phases 
changes to include a smaller or larger fraction of the stars.

\begin{figure}
\includegraphics[angle=0,scale=0.4]{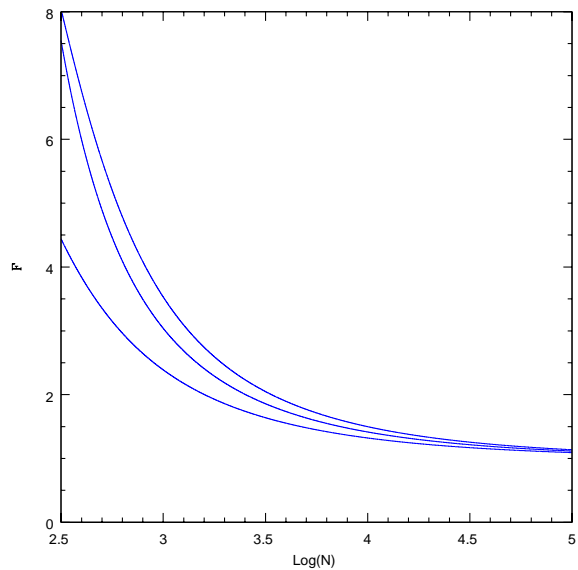}
\caption{Expected factors over which the intrinsic M/L ratios of single stellar populations are expected to vary,
at fixed age and metallicity, as a function of total stellar mass in solar units, for three different ages of
10.5 Gyr, 5 Gyr and 2Gyr.}
\label{fig:picture}
\end{figure}

\section{Conclusions}\label{ccl}

After calculating directly statistical distributions for the inherent $M/L$ ratios of small stellar
populations, we find that in the low mass range of the local ultra-faint dwarf spheroidals, assigning a $M/L$
ratio to a stellar population changes from the deterministic problem of finding the value which corresponds to 
an infinite population having a required metallicity, age and star formation history, to an entirely probabilistic 
situation. Indeed, below total stellar masses of $3 \times 10^{3} M_{\odot}$, the $M/L$ distributions become so 
broad, that the probabilistic nature of the problem becomes the dominant ingredient, relegating age and 
metallicity to a secondary role in establishing the intrinsic $M/L$ ratio of a stellar population. This 
is particularly relevant to the study of the local ultra-faint dwarf spheroidals, as determining their 
baryonic masses through assigning fixed, standard $M/L$ values, can easily lead to significant error.
This is particularly delicate as the $M/L$ distributions which result are far from symmetric about $(M/L)_{\infty}$,
being heavily and systematically skewed towards higher values.

\section*{Acknowledgements}
The author acknowledges the input of an anonymous referee as helpful in reaching a clearer final version, 
and financial support from UNAM-DGAPA grant IN103011-3.

\end{document}